# Step bunching with both directions of the current:

# Vicinal W(110) surfaces versus atomistic scale model

Olzat Toktarbaiuly[1,2,†], Victor Usov[1,†], Cormac Ó Coileáin[1,†], Katarzyna Siewierska[1], Sergey Krasnikov[1], Emma Norton[1], Sergey I. Bozhko[1,3], Valery N. Semenov[1,3], Alexander N. Chaika[1,3], Barry E. Murphy[1], Olaf Lübben[1], Filip Krzyżewski[4], Magdalena A. Załuska-Kotur[4], Anna Krasteva[5], Hristina Popova[6], Vesselin Tonchev[4, 7,*], Igor V. Shvets[1]

[1]Centre for Research on Adaptive Nanostructures and Nanodevices (CRANN), School of Physics, Trinity College Dublin, Dublin 2, Ireland
[2]Nazarbayev University, 53 Kabanbay Batyr Avenue, Astana 010000, Kazakhstan
[3]Institute of Solid State Physics RAS, Chernogolovka, Moscow district 142432, Russian Federation
[4]Institute of Physics, Polish Academy of Sciences, Warsaw, Poland
[5]Institute of Electronics, Bulgarian Academy of Science, Sofia, Bulgaria
[6]Institute of Physical Chemistry, Bulgarian Academy of Science, 1113 Sofia, Bulgaria,
[7] Faculty of Physics, Sofia University, 1164 Sofia, Bulgaria
† These authors contributed equally
* corresponding author: tonchev@phys.uni-sofia.bg

**Abstract.** We report for the first time the observation of bunching of monoatomic steps on vicinal W(110) surfaces induced by step up or step down currents across the steps. Measurements reveal that the size scaling exponent $\gamma$, connecting the maximal slope of a bunch with its height, differs depending on the current direction. We provide a numerical perspective by using an atomistic scale model with a conserved surface flux to mimic experimental conditions, and also for the first time show that there is an interval of parameters in which the vicinal surface is unstable against step bunching for both directions of the adatom drift.

## I. INTRODUCTION

The observation that monoatomic steps on the vicinal surfaces of Si(111) break their initial equidistant distribution under the influence of an electric field to form wide terraces almost free of steps [1], led this system to become the architype for a class of surface instabilities – the bunching of straight steps. This surface in particular is also the most studied both experimentally and theoretically, although step bunching (SB) has been observed in other experimental systems as well [2-14]. However, the step bunching behaviour on Si(111) is unique in part due to the scale of the bunches produced by electromigration alone. One of the main features of this instability on vicinal Si(111) remains a puzzle, the rich reentrant behavior dependent on temperature, with four intervals where alternating directions of the heating current cause SB. In two distinct intervals SB will only occur with step-up (SU) current while in the other two a step-down (SD) current is required. For each of these intervals when the direction of the current required for SB is reversed, de-bunching is observed. The situation is even more complicated when one considers the conditions of growth [15]and equilibrium [16,17]. Theoretical explanations for the instability have been provided by S. Stoyanov for both SD [18] and SU [19] SB regimes during sublimation. The former remains within the Burton-Cabrera-Frank formalism [20] based on the behavior of a single step surrounded by two terraces with uneven contribution to the step velocity. The latter is the result of a collective effect over many step-step distances requiring the additional assumption of *step transparency* – which is to say during their diffusion along the surface, adatoms easily cross surface steps hopping onto the next terrace[19]. There exists no unified theoretical description by which the four temperature intervals of step bunching on Si(111) can be described. In what follows we present experimental results for vicinal surfaces W(110), showing for first time electromigration induced bunching of straight steps for two oppositely oriented currents at the same temperature. Then, an atomistic scale model (vicCA) is used to provide numerical evidence for such a behavior in a model system where the drift direction is reversed while keeping all other parameters within the same ranges.. Experimental and theoretical observations show that bunching can be obtained for effectively the same conditions for both SD and SU drift.

## II. STEP BUNCHING ON W(110) SURFACE

From a technological perspective the patterning of surfaces is used in various experiments for the guided synthesis of nano-objects with reduced dimensionality – nano-dots, nano-wires, etc. [21-26]. As noted the self-assembly of macro-scale steps under the influence of electro-migration is not limited to Si, one of the first observed such systems is probably the surface of tungsten, where systematic studies on the effect of electric fields on tungsten began with lamp filaments [27]. More generally, the effect of electromigration on metal interconnects is still the focus of contemporary research [28]. Soon after the first studies, the patterning of polycrystalline tungsten surfaces by direct currents was reported [29]. An important observation was made by O'Boyle [30] – the direction of the electromigration of tungsten ions is toward the cathode. Based on observations of faceting close to the (110) surfaces of a cylindrical tungsten crystal by Zakurdaev [14], Geguzin and Kaganovski [31] developed the first theoretical perspective on crystal surface instabilities under charge and heat transfer. A modern and systematic treatment of the kinetic faceting of low index surfaces of W was provided by Zhao [32]. Both theory [31] and experiment [32] concluded that only one of the two possible electric field directions should destabilize the low index surfaces. Our experimental results for vicinal surfaces W(110) show electromigration induced step bunching for two oppositely oriented currents at the same temperature.

In order to investigate whether the SB instability can be induced on a metal, high quality W single crystals are used, grown using the floating zone technique with a low density of dislocations and the crystal quality was monitored by X-Ray diffraction (XRD). Rectangular 10×1.5×0.5 mm strips were cut from a crystal ingot at different angles to the W(110) plane with a long side and the miscut direction aligned to the [1-1-2] crystallographic axis. To prepare the vicinal W(110) surfaces prior to DC annealing, the W(110) samples were cleaned and their quality was analyzed in an ultra-high vacuum chamber with scanning tunneling microscopy (STM) capability. To clean the surfaces the vicinal tungsten samples were first annealed using electron bombardment at 1300 °C in an oxygen atmosphere ($p=1\times10^{-6}$ Torr) for 60 minutes to remove the carbon contamination. This was followed by several flash heatings at 2100 °C to remove oxygen from the surface. The cleaning process was repeated until the carbon and oxygen impurities could not be detected by Auger Electron Spectroscopy. The quality of the vicinal surface was finally verified by STM as seen in Fig. 1a, which shows a regular (unbunched) vicinal surface with the characteristic pattern of vicinal W(110). As a final preparation step the samples were oxidized at 1300 °C in an oxygen atmosphere ($p=1\times10^{-6}$ Torr) for 60-90 minutes to create protective oxide layer for transfer from the UHV chamber into the primary experimental setup used to initiate the step bunching. The step bunching process was conducted in separate setup that combines independently controlled direct current and irradiative heating, this was previously used to decouple electric field from temperature in studies of Si(111) [33]. The strips were mounted between the two electrical contacts of the DC annealing chamber's sample holder and inserted into the alumina heating crucible. The procedure began with each sample being outgassed for 24 hours at 700 °C. To remove the protective surface oxide, the heating cell was brought to 1300°C for 1 hour. Next, without radiative heating, a direct current of 5A was applied in the intended primary annealing direction for 24 hours. The electric field was applied perpendicular to the orientation of atomic steps. Finally, maintaining the same current direction, the crucible temperature was raised to 1500 °C and direct currents of $I = 6$ A ($E \approx 0.025$ V/cm), $I = 12$ A ($E \approx 0.05$ V/cm) were applied for 6 hours to initiate the SB process. The annealing temperature of 1500 °C was chosen to increase adatom mobility, this is considerably below the melting point of tungsten (3422°C) and as such no significant evaporation of surface material is expected. Furthermore, due to the low resistance of the samples, joule heating of the sample did not influence the overall temperature. After cooling to room temperature, samples were removed for ex-situ analysis of the step bunched surfaces using atomic force microscopy (AFM)

Annealing without an electric field reveals a regular unbunched surface, however the application of an electric field by means of a current in either the step-up or step-down directions results in a clear step bunched morphology, as shown in Fig. 1b, and 2. This is in sharp contrast to Si(111) where SB instability is only observed for a single current orientation to the offcut direction in each temperature interval. From the AFM images it is clear that there are large straight step bunches, up to 120nm high, separated by wide terraces aligned to the W(110) plane. Wider terraces are decorated by S-shaped crossing steps, similar to those observed for Si(111)[34,35], suggesting the presence of an adatom concentration gradients across the terraces. From cross sections of these bunches we can conclude that the amplitudes of the step bunches produced by SU current direction are different than those for step-down currents. To quantify the morphology, cross sections were taken of the plane leveled bunches and the first derivatives of these profiles were used to determine the maximum slope within the bunches, as shown in Fig. 3. Plotting the height of the bunches against maximum slope, which is determined by the minimum separation of atomic steps within the bunch, reveals distinct scaling relationships depending on the applied current direction, as shown in Fig. 3. The measured values of the size scaling exponent $\gamma$ are 0.59 and 0.67 for step-up and step-down currents respectively, and follow the relationship $l_{min} \sim N^{-\gamma}$. $l_{min}$ is the minimal step-step distance in the bunch, it is inversely proportional to the maximal slope of the bunches (see Figure 3), and the number of steps in the bunch is $N$. These values are similar to those obtained for vicinal Si(111) where SB is only observed for a single current direction in each separate temperature interval (for a recent review see [36]).

The step bunches produced with a step-down field show the slope is always highest somewhere towards the middle of the bunches, as shown in Fig. 3b. Together with the finding that the slope increases when the bunch size increases, this permits to classify the SB process on W(110) as belonging to the B2-type [37]. The difference in the size-scaling exponents, ~2/3 in the SD case and ~ 3/5 in the SU case, hints that there is a change in the SB mechanism. It is suggested that the influence of the electronic wind may be responsible, by eroding the step structure more aggressively when "blowing" up-step for step-down fields. Therefore, the steps become rougher, resulting in a higher density of kinks, and so adopt a non-transparent interactions and are effectively more repulsive. This could explain the quantitative difference observed in Fig. 4, where the slopes obtained with step up field are higher than these obtained with step down field for matching bunch heights (note that the current and thus the electric field in step down direction is twice that in step up in this case). A similar difference was observed in a study of the widths of the step bunches in "equilibrium" conditions on Si(111) [16]. The bunches formed by SD currents were larger, while those produced by SU currents were more compressed, and thus displayed higher slopes.

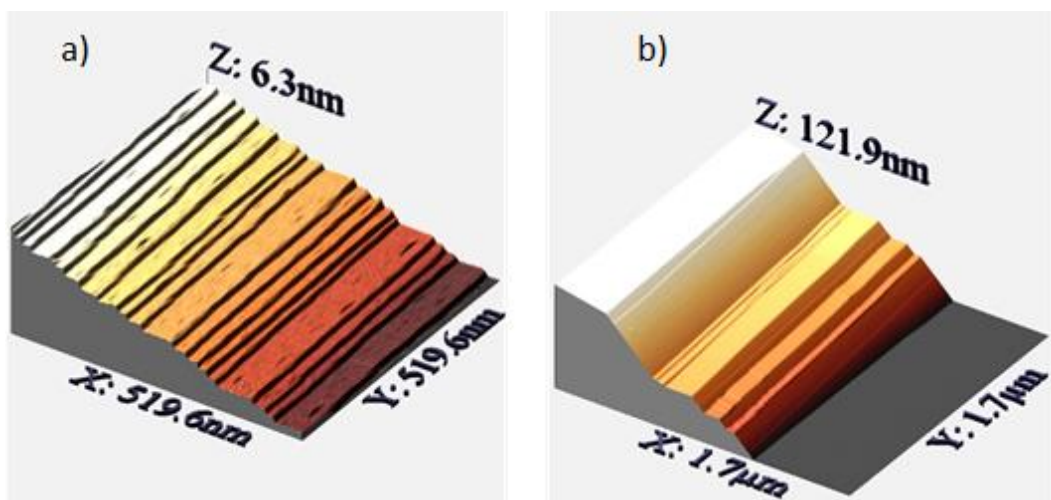

**Figure 1.** (a) STM images of flat unbunched surface of W(110) with a miscut of 3.2° after cleaning procedure (b) STM image of bunched W(110) surface with a miscut of 2.6° annealed with step-down current of 12A DC for 6 hours

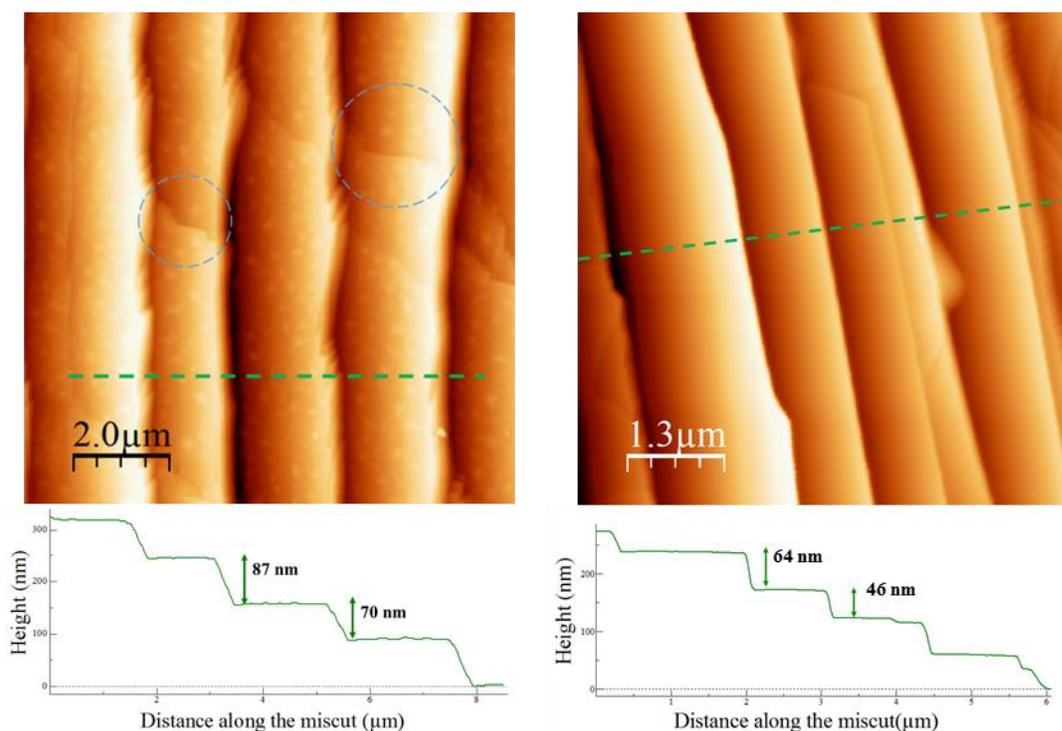

**Figure 2.** (a) AFM of W(110) step-bunched at 1500 °C with assistance of the step down 12 A DC-current passed in [1-1-2] direction for 6 hours (b) AFM of W(110) step-bunched at 1500 °C with the step-up 6 A DC-current passed in [-112] for 6 hours. Crossing steps are highlighted. A cross sectional profiles of locally plane-fitted surfaces are shown below.

| Step-up regime of SB (SU) | Step-down regime of SB (SD) |
|---|---|
| 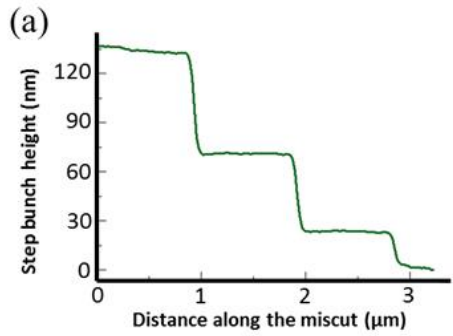 | 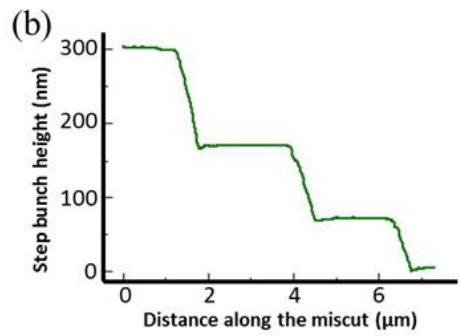 |
| 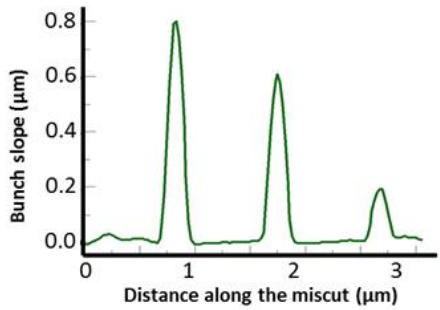 | 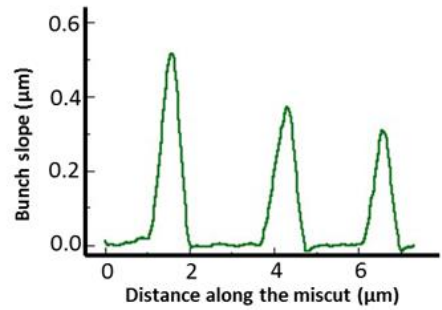 |

Figure 3 Profile of typical step bunched surfaces and the slopes attributed to the step bunches. Above plane levelled line profile along samples with an off-cut by 2.6°, and below the corresponding absolute value of the slope. For a surface annealed with a current of (a) 6A in the step-up direction and (b) 12A in the step-down direction.

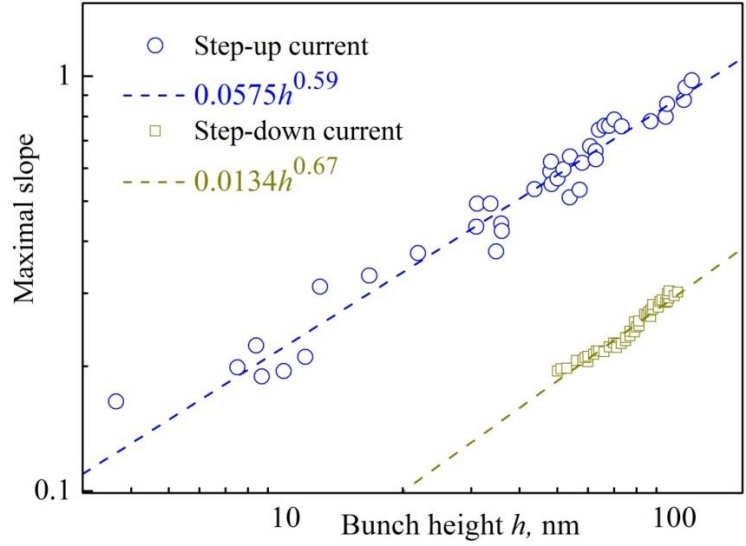

**Figure 4**. Scaling relationship between the maximum slope and height of step bunches on W(110) surface with miscut 2.6° annealed by step-up direct current of 6 A and compared with surface with a miscut of 2.6°, annealed with step-down 12 A current. The exponents obtained, 0.67±0.02 and 0.59±0.02, are distinguishable based on the error of the fit.

## III. THE MODEL

A simple model is proposed that shows step bunching for both drift directions for the same parameters. Our 1D model of a vicinal crystal surface with adatoms on it is a *conserved* version of the recently introduced atomistic scale model[38,39], to emulate the lack of evaporation on the tungsten surface at 1500°C. The vicinal stairway descends from left to the right and initially the steps have an equidistant separation $l_0$. Thus, the length of the system is an integer multiple of $l_0$ with periodic boundary conditions. On top of this surface there is a layer of adatoms. The system dynamics are a combination of Cellular Automaton (CA) and Monte Carlo (MC) simulation techniques. To simulate adatom diffusion the MC method is applied. The coordinates of the adatoms are stored in a separate array. Adatom diffusion along the vicinal surface is influenced by a unidirectional bias δ, a number that modifies the adatom hopping probability along the surface, with 1/2±δ in the right/left direction. Negative values of δ represent a bias in the step up direction, while a positive δ makes the step down direction preferable. During a single diffusional step all the adatoms make on average $n_{DS}$ hop attempts towards neighboring lattice sites with a modified probability of 1/2±δ, and if the chosen site is not occupied already the hop is executed. The collective effect of the biased diffusion of the adatom population is thus the *drift*. The crystal growth aspect of the steps is simulated using the CA module. It operates according to a simple predefined rule – adatoms right of the step attach to it. The conditionality of attachment can be modified according to a predefined probability $p_G$<1. When an adatom attaches to a step, it is deleted from the array of the adatoms and the surface height at its position is increased by 1, causing an effective lateral step to move one lattice position to the right. A detailed description of the above process is available in [38,39]. However, we extend the model here by adding the possibility of adatom-from-surface detachment. This process is the opposite of crystal growth and causes step motion to the left. In such a way we can simulate both crystal growth and sublimation. The sublimation is realized as follows - every site where a single or macro step is present is checked to determine if it is at the surface or whether there is an above adatom. When free, detachment from the step is executed with a detachment probability $p_S$ and if this occurs the height of the lattice at that site is reduced by 1, a new adatom is added to the adatom array at that place. The particle concentration sets itself up as a result of the balance between the attachment and detachment probabilities $p_G$ and $p_S$. This equilibrium value is around $c_0=p_s/(p_s+p_G)$. Note, that the model in this form is conserved (see also [40,41]), which means that it describes the system annealed without any flux of incoming or outgoing particles like the experimental system described above. Furthermore this better describes vicinal W(110), which is unlike the experimentally realized and theoretically modeled "equilibrium" state of Si(111)[16] where the intensive fluxes from and to the surface exist but are fully compensated.

Thus a single *time step* increment of the model procedure consists of: (i) parallel attachment of adatoms, nearest neighbors of the steps from the lower terrace; (ii) $n_{DS}/2$ serial diffusional hops per adatom, (iii) parallel detachment from the steps with the probability $p_S$ but only when no adatom is present above the chosen step; and (iv) $n_{DS}/2$ serial diffusional steps. The parallel execution of (i) and (iii) is achieved by postponing the decision for each "eligible" adatom in a mirror array in a Cellular Automaton fashion. The multiple repetitions of the procedure above, lead to the equilibration of adatom concentration around a value $c_0$. Due to the allocation of the adatoms in a separate array, during the diffusional hops they do not "feel" the surface steps. Thus, with increasing $n_{DS}$ we not only depart from a diffusion-limited regime ($n_{DS}$=1) towards an attachment-detachment limited regime but simultaneously we increase the manifestation of *step transparency*. Depending on the parameters set, a variety different dynamic behaviors are observed - from stable equidistant step trains, through the formation of irregular structures to well defined, regular bunches. It should be noted that although our model does not incorporate step-step repulsions in the vicCA, we show in a parallel study[42] that it reproduces the time-scaling of the bunch size *N* derived from models where the step-step repulsion is taken into account (BCF-type models). Thus, while the time-scaling of the bunch size, $N = 2\sqrt{T/3}$, where $T$ is the properly re-scaled time[43], is universal both in terms of scaling exponent and numerical pre-factor, it is the behavior of the bunch width/macrostep height that is model dependent. Studying models without the stabilizing role of the step-step repulsions is not unusual – two of the most influential studies in the field of surface instabilities [18,44] are also focused on the destabilizing factors only – electromigration of the adatoms and uneven coefficients for step attachment from the two terraces neighboring a step, respectively. The conclusions of such studies could be also relevant for explaining experiments in case the step-step repulsions in the experimental system are negligible.

III.1. Stability analysis

We analyze the stability of the defined model against step bunching. A specific approach is required due to the lack of analytical expressions, such as these used for example in [45]. Using the established monitoring schemes we probe the surface stability by following the systems' evolution with time $t$ and plot "stability diagrams" in the space ($p_1$, $p_2$, $N$), where $p_1$ and $p_2$ are the chosen model parameters and $N$ is the average bunch size. The "stability diagrams" sketch the surface stability against step bunching with both directions of *drift*. First, we find the most favorable values of the initial vicinal distance $l_0$ in Fig. 4a. This plot shows that SB is somewhat preferential depending on the drift direction, for large values of $l_0$ favor SB with SU drift, SB with SD drift is favored by moderate values of $l_0$. Still, there is an overlap of the two regions with respect to optimal $l_0$ – values of $l_0$ around 10 are most favorable for SB with both drift directions. Another interesting cross-section of our stability analysis is presented in Fig. 5b where it can be seen that the surface stability depends on the detachment probability $p_s$. While the values of $l_0$ can be controlled experimentally by choosing the miscut angle, $p_s$ could be thought of as a parameter that measures the thermal activation of the adatoms. Thus Fig. 5b reveals an interesting asymmetry – while the occurrence of SB with SD drift directions is almost independent of $p_s$ in order to observe SB with a SU drift direction large values of $p_s$ are needed. Note that large $p_S$ values mean a high probability of desorption from the steps, which can be attributed to relatively high temperatures experimentally. Even for such values there is a stability gap spanning the small values of $|\delta|$. Whether this gap shrinks for longer simulation times is a matter of further studies, but it is not expected that these longer times will decrease the values of $N$ in regions where the instability has already developed.

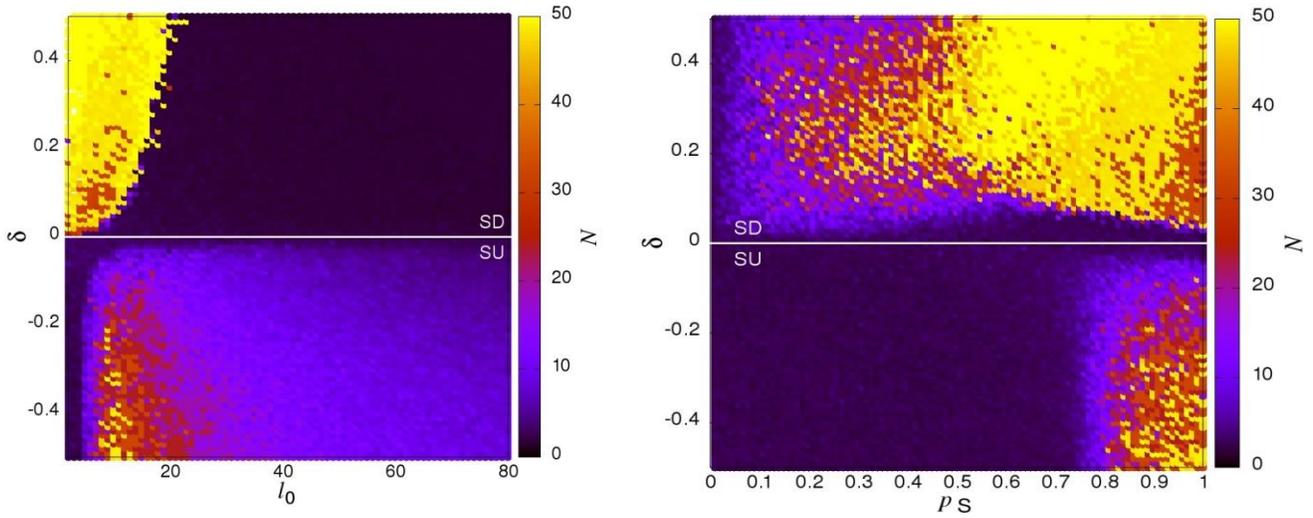

Figure 5. Surface stability as probed using the average bunch size $N$. The values of the parameters are scanned on a dense mesh. a) in the plane ($\delta$, $l_0$); $n_{DS}$=10; $p_S$ =0.9 b) in the plane ($p_S$, $\delta$); $l_0$=10; $n_{DS}$=10

An interesting perspective on the peculiarities of the surface stability is provided by Fig. 6. For SU drift, Fig. 6a within the instability region, $p_S > 0.75$, there is a sharp stability transition with $p_S$ and a gradual increase in the stability along $l_0$. When the drift is SD, Fig. 5 b, there is practically no step bunching for values of $l_0$ greater than 30 while for values of $p_S < 0.7$ there is a gradual increase of the stability with decreasing $p_S$. Therefore, one can outline a region in the ($p_S$, $l_0$)-plane where SB occurs for both SU and SD drifts and two other regions where it is found with only one of the drift directions. In Fig. 7 we show a plot in the plane ($\delta$, $n_{DS}$). Similarly, as for the previous two cases, there is a stability gap between the zones of instability spanning small values of $|\delta|$. Whether this gap shrinks for longer times of simulation is again a matter of further studies. Again, as with previous cases bunching is evident and is particularly strong for the SD bias, and weaker for the SU bias, being concentrated only around small values of diffusion steps.

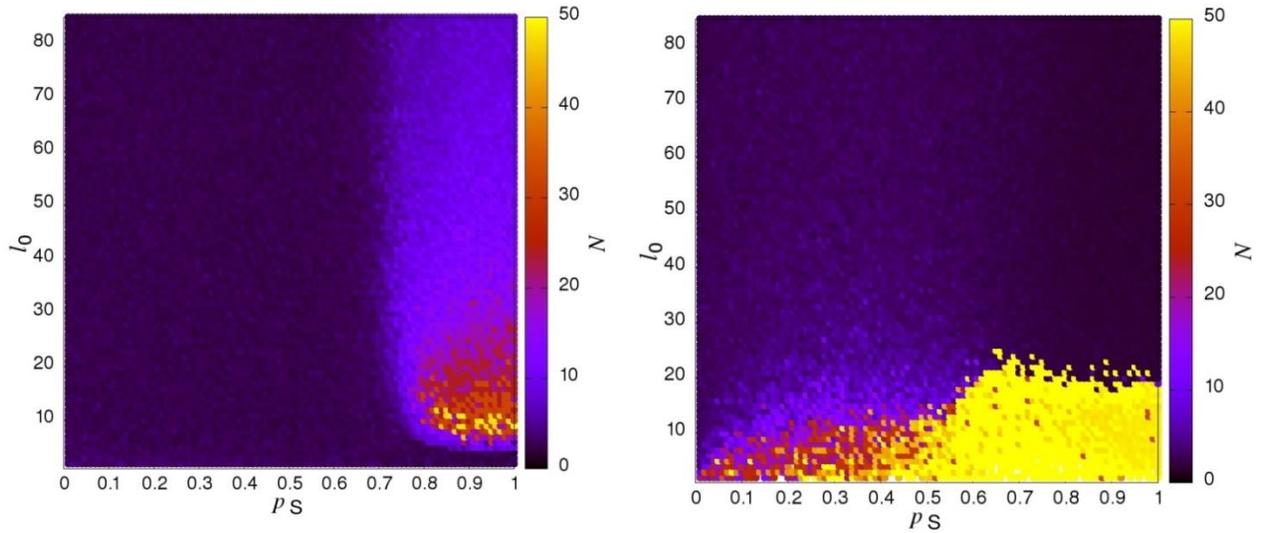

Figure 6 a. Surface stability for SU drift, $\delta = -0.4$, in the plane ($p_S$, $l_0$). $n_{DS}=10$, b. Surface stability for SD drift, $\delta = 0.4$ in the plane ($p_S$, $l_0$). $n_{DS}=10$;

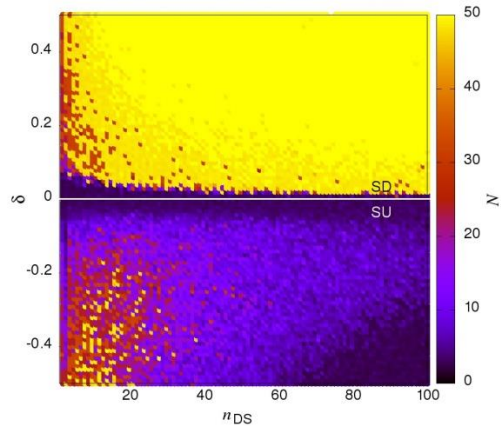

Fig. 7 Stability diagram in the ($\delta$, $n_{DS}$)-plane in terms of bunch size $N$ as function of the model parameters, $p_S = 0.8$ and $l_0=10$;.The resulting values of the bunch size are obtained in each of the points in for the same time $t=500000$.

A complementary plot in plane ($l_0$, $n_{DS}$) is shown in Fig. 8 for both the SD and SU biases. It can be seen that step bunching occurs for all $n_{DS}$ values examined, but for narrow terraces only, i.e. for $l_0$ smaller than 20. SU bunching is weaker, visible for terraces greater than 5, but diminishes if too large, and is more visible for smaller values of diffusion steps. It should be noted that moderate values of $l_0$ and $n_{DS}$ around 10 are favorable for SB for both drift directions.

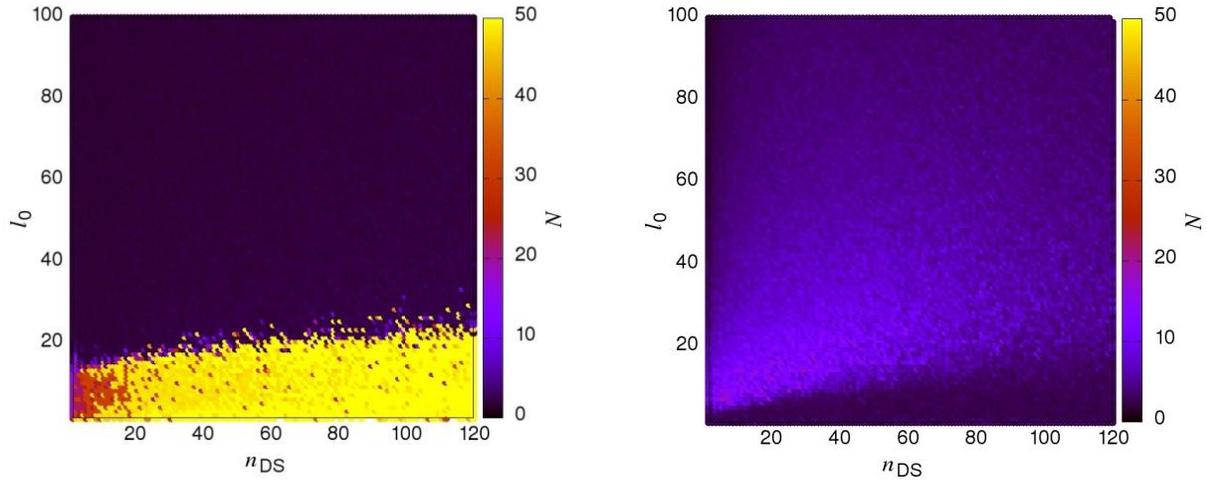

Fig 8 a. Surface stability as a function of initial terrace with $l_0$ and number of diffusional steps $n_{DS}$ for SD bias, $\delta=0.4$ in left panel and for SU bias, $\delta=-0.4$ in right panel. Both systems are studied for $p_G=1$ and $p_S=0.9$.

Let us also analyze the ($p_S$, $n_{DS}$) plane. Fig. 9a is obtained for step-down direction of drift - the SB phenomenon is most intensive for values of $p_S$ above 0.5 for all values of $l_0$ while Fig. 9b examines for step-up drift. This last plot is surprising - it shows two separate regions of instability: for high $p_S$ and low $n_{DS}$ and for low $p_S$ and high $n_{DS}$. This first instability overlaps with the large instability present in the SD bias diagram, and is the one we propose as a model for the bunching phenomenon observed experimentally on the W(110) surface. The second area of instability in the SU has no corresponding area in SD stability diagram, but exists for low $p_S$ data, and can be responsible for system behavior under different experimental conditions than those studied here. Conceptually similar stability analyses were published recently of Monte Carlo models where the source of the instability is Erhlich-Schwoebel barrier[46,47].

III.2. Step trajectories

It can be seen in Figs. 5-9 that SD drift causes bunching for a broad variety of parameters. However, the time evolution of the bunches produced by each of the two drift directions is completely different as shown in Fig. 10. In left panel trajectories of multisteps and steps are plotted for the case of a SD bias. It can be seen that bunches slowly move in the up step direction in the process of step exchange. Bunches grow absorbing more and more of individual steps that come from other, vanishing bunches. Such evolution of bunches can be compared with trajectories plotted in right panel for SU case. It is shown that pairs of bunches are slowly drawn together and then merge to create single larger bunches, and notably do not exchange steps.

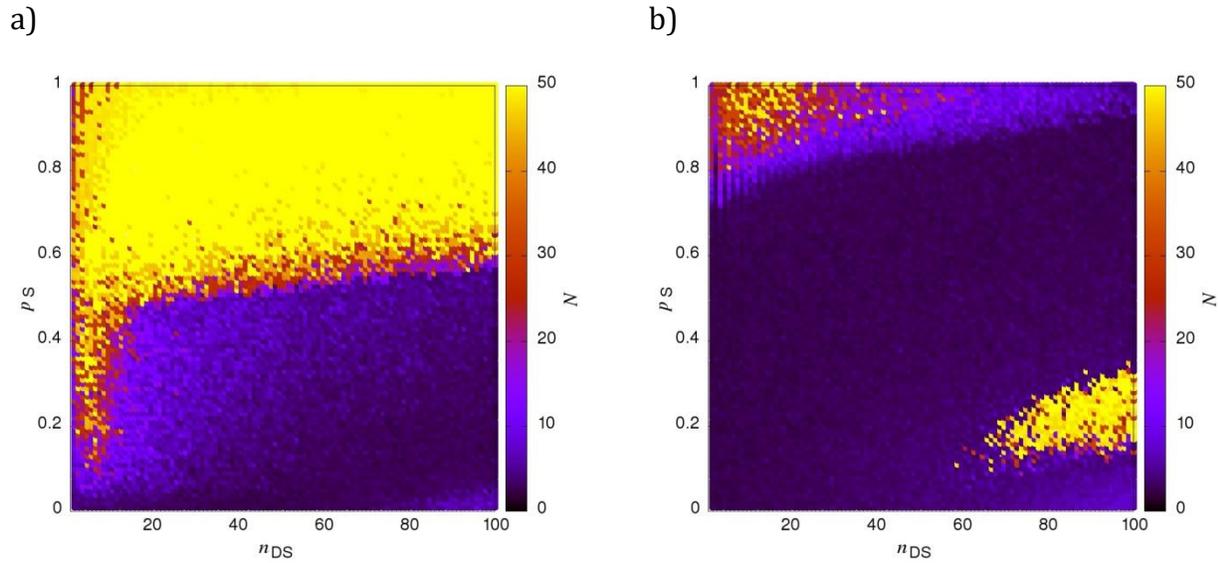

Fig. 9 Stability diagrams in the ($p_S$, $n_{DS}$)-plane in terms of bunch size $N$ as function of the model parameters and $l_0$=10; a) $\delta$=0.4 and b) $\delta$=-0.4. The resulting values of the bunch size are obtained in each of the points in for the same time $t$=500000.

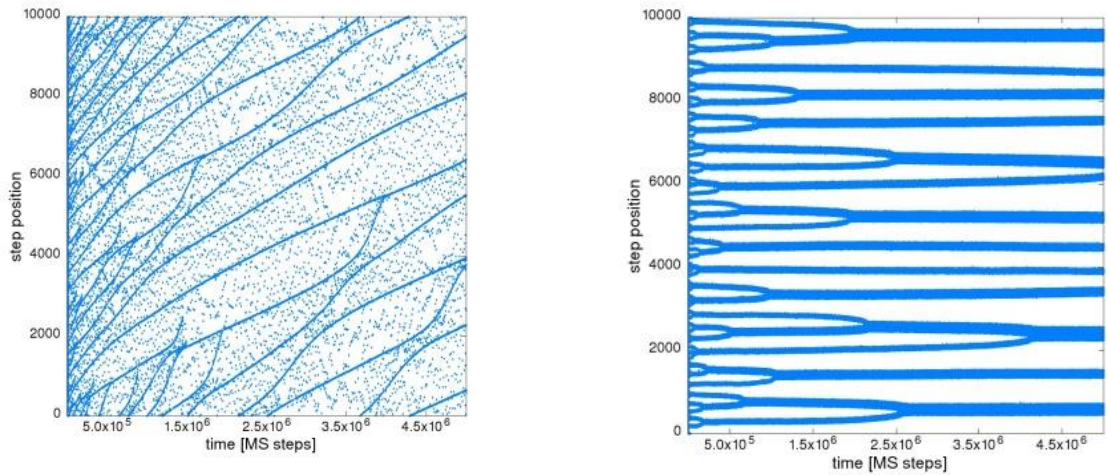

Figure 10. Step positions as a function of number of simulation steps. Trajectories are calculated for $l_0$=10, $p_G$=1, $p_S$=0.9, $n_{DS}$=10, and $\delta$=0.4 (SD) for panel at left hand side and $\delta$=-0.4 (SU) for right hand side panel.

## IV. DISCUSSION AND CONCLUSIONS

We have shown that two oppositely oriented electric fields cause step bunching on vicinal W(110) surfaces annealed at the same temperature of 1500°C. The shape of the resulting bunches was illustrated and analyzed by comparison of the maximum slope vs. bunch size dependence (scaling) in both cases. This is the first report of step bunching when the current is applied in both directions and when all terraces are equivalent – the experimental observations [48] of SB on Si(001) and theoretical modelling [49,50] reveal the importance in this system of the alternation of two different types of terraces with respect to the diffusion of adatoms on them – slow or fast. We provide a possible hypothesis to explain qualitatively the observations, and to make them consistent with the paradigm based on studies of Si(111) vicinal surfaces, where the step-up field causes step bunching when the steps are transparent and the step down field requires non-transparent steps to cause step bunching. We can suppose for W(110) that when the field direction is oriented step down the *electron wind* is oriented step up, and this roughens the step edges on atomic scale turning them non-transparent. For a step down *electron wind* the steps are not affected and they remain transparent, thus enabling step bunching with step up direction of the field. In our study we go beyond this simple hypothesis and, using a simple effectively-constructed atomistic scale model of step dynamics we find model parameters that reproduce the instability with both directions of the adatom drift. In our model all terraces are equivalent. The character of the process of step bunching depends on the current direction, which is illustrated both by the different scaling behaviors from the experimental data and different shapes of step trajectories in the model. We believe that this example is the first step towards a deeper understanding of step bunch instabilities, and a route to better control of step bunching on the surfaces of other materials.

## ACKNOWLEDGEMENTS

This work is financially supported by Science Foundation Ireland, under Grant No. 12/IA/1264 and Marie Curie IIF grant within the 7-th European Community Framework Programme. OT acknowledges support of the government of Kazakhstan under the Bolashak program and partially funded under the program №0115PK03029 from the Ministry of Education and Science of the Republic of Kazakhstan. KS acknowledges funding through Irish Research Council. The modeling is carried within a BAS-PAS collaborative project. VT, HP and AK acknowledge financial support from Bulgarian NSF T02-8/121214. FK is supported by NCN of Poland, grant 2013/11/D/ST3/02700. VT acknowledges very stimulating working conditions in IP-PAS. Part of the calculations was done on HPC Nestum (BG161PO003-1.2.05). The authors acknowledge that the first observation of the electromigration driven step bunching on W(110) was made by Victor Usov and Cormac Ó Coileáin. VT is thankful to D. Rogilo and E. Rodyakina (Novosibirsk) for pointing at the works studying SB on Si (001).